\documentstyle[twoside,fleqn,col2,epsfig]{article}
\newcommand{\be}{\begin{eqnarray}}
\newcommand{\ee}{\end{eqnarray}}
\newcommand{\dslash}{\partial \hskip -0.5em /}

\newcommand{\AmS}{{\protect\the\textfont2
  A\kern-.1667em\lower.5ex\hbox{M}\kern-.125emS}}

\title{Nucleon structure functions from a chiral 
soliton\thanks{Talk presented by HW at the Int. Conf. QCD--98, 
Montpellier, July 1998. 
Work supported by DFG (Re 856/2--3, We 1254/3-1)
and DOE (DE--FE--02--95ER40923).}
}

\author{H. Weigel\address{Institute for Theoretical Physics,
        T\"ubingen University\\
        Auf der Morgenstelle 14, D--72076 T\"ubingen, Germany}%
        \thanks{Heisenberg--Fellow}%
        \thanks{Address after Sept.~$1^{\rm st}$~1998: 
         Center for Theoretical Physics,
         Laboratory for Nuclear Science and Dept. of Physics,
         Massachusetts Institute of Technology,
         Cambridge, Mass. 02139.},
        L. Gamberg\address{Department of Physics and Astronomy,
        University of Oklahoma\\
        440 W. Brooks Ave, Norman, OK 73019--0225, USA},
        H. Reinhardt$^{a}$ and O. Schr\"oder$^{a}$
       }
       
\begin{document}

\begin{abstract}
We study nucleon structure functions within the bosonized version of
the Nambu--Jona--Lasinio (NJL) model in which the nucleon emerges as 
the soliton in the chiral field.   Upon boosting to the infinite momentum 
frame and performing the $q^2$--evolution in the context of the Gottfried 
sum rule for electron nucleon scattering we determine the intrinsic scale 
$\mu^2$ of the NJL chiral soliton. We also compute the leading twist 
contributions of the polarized structure functions $g_1$ and $g_2$.
We compare these model predictions with experiment by evolving them 
from $\mu^2$ to the scale where the data are taken. Analogously we 
analyze the chiral--odd structure functions $h_T$ and $h_L$. Finally 
we generalize the treatment to flavor SU(3).
\end{abstract}

\maketitle

\section{INTRODUCTION}

The purpose of this investigation is to provide a link between two 
successful although seemingly unrelated pictures of baryons. On one 
side we have the quark parton model which describes the scaling 
behavior of the structure functions in deep inelastic scattering (DIS) 
processes. The deviations from these scaling laws are computable in the 
framework of perturbative QCD. On the other side we have the chiral 
soliton approach motivated by generalizing QCD to an arbitrary number 
of color degrees of freedom, $N_C$. For $N_C\to\infty$, QCD is equivalent 
to an effective meson theory. Although this theory 
is not explicitly known it can be modeled by assuming that at low 
energies only the light mesons (pions, kaons, $\rho$, $\omega$) are 
relevant. The major building block to model the effective theory
is chiral symmetry and its spontaneous breaking. Baryons  
emerge as non--perturbative (topological) configurations of the 
meson fields, the so--called solitons. The link between these two 
pictures can be established by computing structure functions within 
a chiral soliton model for the nucleon from the hadronic tensor  
\be
\vspace{-0.1cm}
W^{ab}_{\mu\nu}(q)&=&\frac{1}{4\pi}\int d^4 \xi \,
{\rm e}^{iq\cdot\xi} \nonumber \\ &&
\hspace{-0.5cm}\times
\langle N(P) |\left[J^a_\mu(\xi),J^{b{\dag}}_\nu(0)\right]|N(P)\rangle \ ,
\label{deften}
\ee
which describes the strong interaction part of the DIS 
cross--section. In eq (\ref{deften}) $|N(P)\rangle$ refers to the 
nucleon state with momentum $P$ and $J^a_\mu(\xi)$ to the hadronic 
current suitable for the process under consideration. In most
soliton models the current commutator (\ref{deften}) remains 
intractable. However, the Nambu and Jona--Lasinio (NJL) model 
\cite{Na61} of quark flavor dynamics contains simple current 
operators since derivatives of the quarks fields only appear in form of 
a free Dirac Lagrangian. Most importantly, the bosonized \cite{Eb86}
version of the NJL--model contains soliton solutions \cite{Al96,Ch96}. 
Here we confine ourselves to presenting the key issues and results of 
the structure function calculation; details may be traced from 
our recent papers \cite{We96,We97a,We97b,Sch98}. For subsequent 
studies see also \cite{Di96}.

\vspace{-0.2cm}
\section{THE NUCLEON IN THE NJL CHIRAL SOLITON MODEL}
\vspace{-0.2cm}

The NJL--model under consideration contains a chirally symmetric quartic 
quark interaction in the scalar--pseudoscalar channel. After bosonization 
the action reads \cite{Eb86}
\be
\vspace{-0.1cm}
{\cal A}={\rm Tr}\, {\rm ln}_\Lambda
\left(i\dslash\hspace{-1pt}-\hspace{-1pt} m U^{\gamma_5}\right)
\hspace{-1pt}+\frac{m_0m}{4G}{\rm tr}\left(U\hspace{-2pt}+\hspace{-2pt}
U^{\dag}\hspace{-2pt}-\hspace{-2pt}2\right) . \hspace{-5pt}
\label{bosact}
\ee
The pion fields 
$\mbox{\boldmath $\pi$}$ are contained in the chiral field
$U={\rm exp}(i\mbox{\boldmath $\tau$}\cdot
\mbox{\boldmath $\pi$}/f_\pi)$. In eq (\ref{bosact}) ${\rm tr}$ 
denotes discrete flavor trace while ${\rm Tr}$ also includes
the functional trace. The parameters of the model are the coupling 
constant $G$, the current quark mass $m_0$ and the UV cut--off 
$\Lambda$. These are adjusted to the pion mass $m_\pi=135{\rm MeV}$ and
decay constant $f_\pi=93{\rm MeV}$. This leaves one parameter 
undetermined which we express in terms of the constituent quark 
mass $m$. It arises as the solution to the Schwinger--Dyson (gap) 
equation and characterizes the spontaneous breaking of chiral symmetry. 

An energy functional for non--perturbative, static field 
configurations $U(\mbox{\boldmath $r$})$ is derived from (\ref{bosact}). 
It can be expressed as a regularized sum of single quark energies 
$\epsilon_\mu$ of the associated one--particle Dirac Hamiltonian in 
the background of $U(\mbox{\boldmath $r$})$. The distinct level (v), 
which is strongly bound is referred to as the valence quark state. Its 
explicit occupation guarantees unit baryon number. The classical soliton
$U_{\rm cl}(\mbox{\boldmath $r$})$ is determined by self--consistently 
minimizing the energy functional. To generate states with nucleon quantum 
numbers the (unknown) time dependent field configuration is approximated 
by elevating the zero modes to time dependent collective coordinates
\be
\vspace{-5mm}
U(\mbox{\boldmath $r$},t)=A(t)U_{\rm cl}(\mbox{\boldmath $r$})A^{\dag}(t), 
\ A(t)\in {\rm SU}(2)\, .
\label{collco}
\vspace{-5mm}
\ee
Upon canonical quantization the angular velocities,
$\mbox{\boldmath $\Omega$}=-i{\rm tr}
(\mbox{\boldmath $\tau$}A^{\dag}\dot A)$, are replaced by the 
spin operator $\mbox{\boldmath $J$}$ via 
$\mbox{\boldmath $\Omega$}=\mbox{\boldmath $J$}/\alpha^2$
with $\alpha^2$ being the moment of inertia.
To compute nucleon properties the action (\ref{bosact}) is expanded 
in powers of $\mbox{\boldmath $\Omega$}$ corresponding to an
expansion in $1/N_C$. In particular the valence quark wave--function
$\Psi_{\rm v}(\mbox{\boldmath $x$})$ acquires a linear correction 
\be
\vspace{-5mm}
\Psi_{\rm v}(\mbox{\boldmath $x$},t)&=&
{\rm e}^{-i\epsilon_{\rm v}t}A(t)
\psi_{\rm v}(\mbox{\boldmath $x$})
\label{valrot}\\ && \hspace{-53pt}
={\rm e}^{-i\epsilon_{\rm v}t}A(t)
\left\{\hspace{-1pt}\Psi_{\rm v}(\mbox{\boldmath $x$})
\hspace{-1pt}+\hspace{-1pt}\sum_{\mu\ne{\rm v}}\hspace{-1pt}
\Psi_\mu(\mbox{\boldmath $x$})\hspace{-1pt}
\frac{\langle \mu |\mbox{\boldmath $\tau$}\cdot
\mbox{\boldmath $\Omega$}|{\rm v}\rangle}
{2(\epsilon_{\rm v}-\epsilon_\mu)}\right\}.
\nonumber
\vspace{-5mm}
\ee
Here $\psi_{\rm v}(\mbox{\boldmath $x$})$ refers to the spatial part
of the body--fixed valence quark wave--function with the rotational
corrections included.

\vspace{-0.2cm}
\section{STRUCTURE FUNCTIONS FROM THE SOLITON}
\vspace{-0.2cm}

In order to extract the leading twist contributions to the structure
function one computes the hadronic tensor in the Bjorken limit
\be
\vspace{-0.1cm}
q_0&=&|\mbox{\boldmath $q$}| - M_N x \quad
{\rm with}  \quad
|\mbox{\boldmath $q$}|\rightarrow \infty\, ,
\nonumber \\
x&=&{-q^2}/{2P\cdot q}\quad  {\rm fixed}\ .
\label{bjlimit}
\ee
For localized field configurations, such as the soliton,
the symmetric part of hadronic tensor (which contains the 
unpolarized structure functions) then reads \cite{Ja75}, 
\be
W^{lm}_{\{\mu\nu\}}(q)\hspace{-2pt}&=&
\hspace{-2pt}\zeta\int \frac{d^4k}{(2\pi)^4} \
S_{\mu\rho\nu\sigma}\ k^\rho\
{\rm sgn}\left(k_0\right) \, \delta\left(k^2\right)
\nonumber \\ && \hspace{-2.0cm}
\times\int dt \ {\rm e}^{i(k_0+q_0)t}
\hspace{-2pt}\int d^3x_1\hspace{-2pt} \int d^3x_2 \
{\rm e}^{-i(\mbox{\scriptsize\boldmath $k$}
+\mbox{\scriptsize\boldmath $q$})\cdot
(\mbox{\scriptsize\boldmath $x$}_1-\mbox{\scriptsize\boldmath $x$}_2)}
\nonumber \\ && \hspace{-2.0cm}
\times \langle N |\Big\{
{\hat{\bar \Psi}}(\mbox{\boldmath $x$}_1,t)t_l t_m\gamma^\sigma
{\hat\Psi}(\mbox{\boldmath $x$}_2,0)
\nonumber \\ && 
-{\hat {\bar \Psi}}(\mbox{\boldmath $x$}_2,0)t_m t_l\gamma^\sigma
{\hat\Psi}(\mbox{\boldmath $x$}_1,t)\Big\}| N \rangle .
\label{stpnt}
\ee
Note that the quark spinors are functionals of the soliton.
Here $S_{\mu\rho\nu\sigma}=g_{\mu\rho}g_{\nu\sigma}
+g_{\mu\sigma}g_{\nu\rho}-g_{\mu\nu}g_{\rho\sigma}$ and
$\zeta=1(2)$ for the structure functions associated with the
vector (weak) current and $t_m$ is a suitable isospin 
matrix. The matrix element between the nucleon states 
($|N\rangle$) is taken in the space of the collective coordinates. 

\begin{figure}[t]
\hspace{0.5cm}
\epsfig{figure=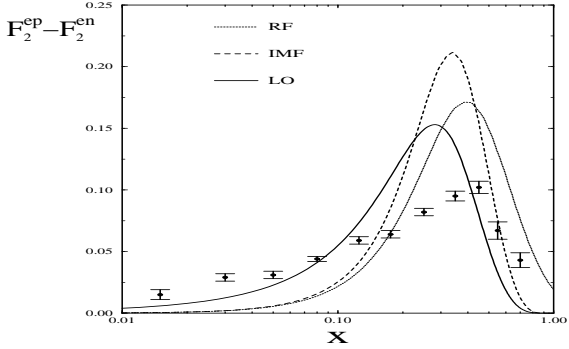,height=4.0cm,width=7.0cm}
~
\vspace{-0.6cm}
\caption{\label{fig_2}The unpolarized structure function
entering the Gottfried sum rule. RF: rest frame, IMF: boosted
to the infinite momentum frame, LO: leading order QCD 
evolution to $Q^2=4{\rm GeV^2}$.}
\vspace{-.2cm}
\end{figure}

The valence quark approximation ignores the vacuum polarization in 
(\ref{stpnt}), {\it i.e.} the quark field operator ${\hat\Psi}$ is 
substituted by the valence quark contribution (\ref{valrot}). For 
constituent quark mass under consideration $m=400{\rm MeV}$ this 
approximation is well justified since this level provides the dominant 
share to static observables \cite{Al96,Ch96}. The structure function 
$F_2(x)$ is obtained from (\ref{stpnt}) by an appropriate 
projection\footnote{In 
the Bjorken limit the Callan--Gross relation $F_2(x)=2x F_1(x)$ is 
satisfied.}. Then eq (\ref{stpnt}) yields the structure functions in 
the nucleon rest frame (RF). In order to obtain proper support 
these are transformed to the infinite momentum frame (IMF)
\cite{Ja80,Ga97}:
\be
\vspace{-2mm}
f_{\rm IMF}=\frac{1}{1-x}\, 
f_{\rm RF}\left(-{\rm ln}(1-x)\right) \, .
\label{rf2imf}
\vspace{-2mm}
\ee
Adopting the point of view that the model approximates QCD at the
low scale $\mu^2$ we apply a leading order DGLAP evolution to 
$f_{\rm IMF}$. Demanding a {\it best agreement} with the data at 
the experimental scale $Q^2$ for the linear combination entering 
the Gottfried sum rule $\left(F_2^{ep}-F_2^{en}\right)$
determines $\mu^2=0.4{\rm GeV}^2$. In figure \ref{fig_2} the resulting
structure function is compared to the data \cite{Ar94}. 
Apparently the gross features are reproduced.  
Moreover the integral of the Gottfried sum rule
\be
S_G&=&
\int_0^\infty \frac{dx}{x}
\left(F_2^{ep}-F_2^{en}\right) 
\nonumber \\
&=&\cases{0.29\ , \ m=400 {\rm MeV}\cr
0.27\ , \ m=450 {\rm MeV}}
\label{gottrule}
\ee
agrees reasonably well with the empirical value $S_G=0.235\pm0.026$
\cite{Ar94}. In particular the deviation from the na{\"\i}ve value 
(1/3) \cite{Go67} is in the direction demanded by experiment. 

\begin{figure}[ht]
\epsfig{figure=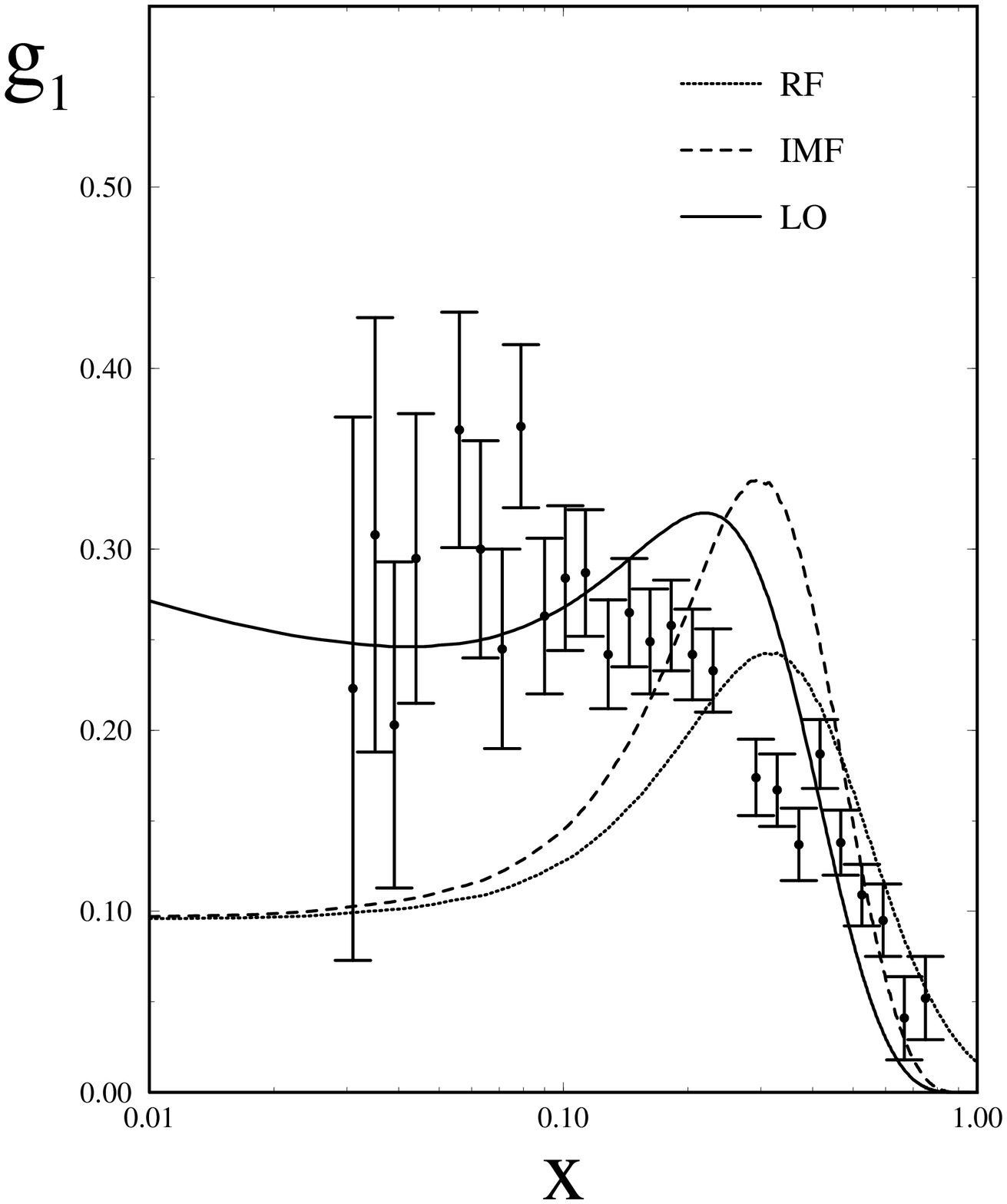,height=4.0cm,width=7.5cm}
~\vspace{0.3cm}\\
\epsfig{figure=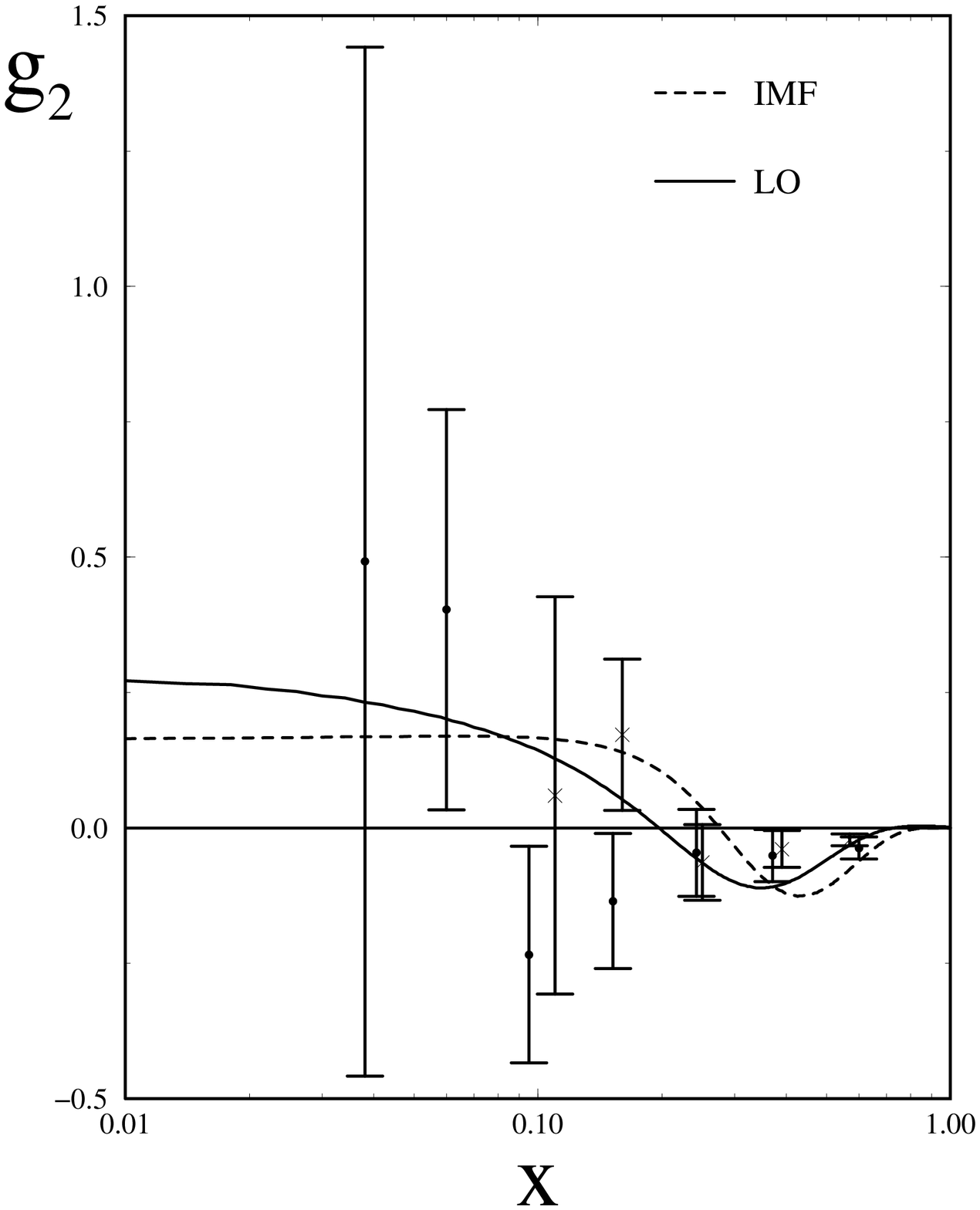,height=4.0cm,width=7.5cm}
\vspace{-0.4cm}
\caption{\label{fig_4}The polarized structure functions $g_1$
(top) and $g_2$ (bottom) after projection and QCD evolution. 
Data from SLAC \protect\cite{slac96}.}
\vspace{-0.3cm}
\end{figure}

\begin{figure}[ht]
\epsfig{figure=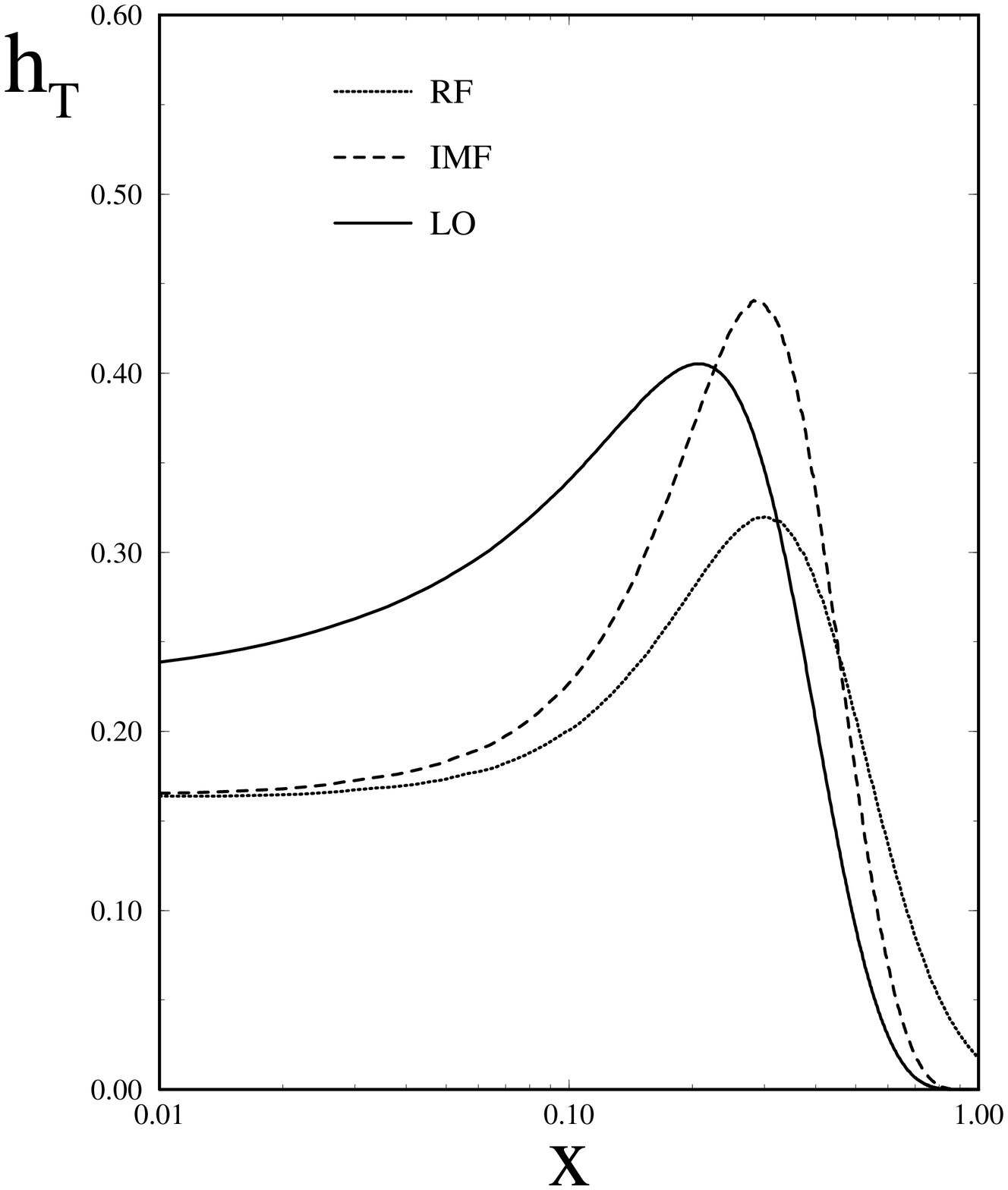,height=4.0cm,width=7.5cm}
~\vspace{0.3cm}\\
\epsfig{figure=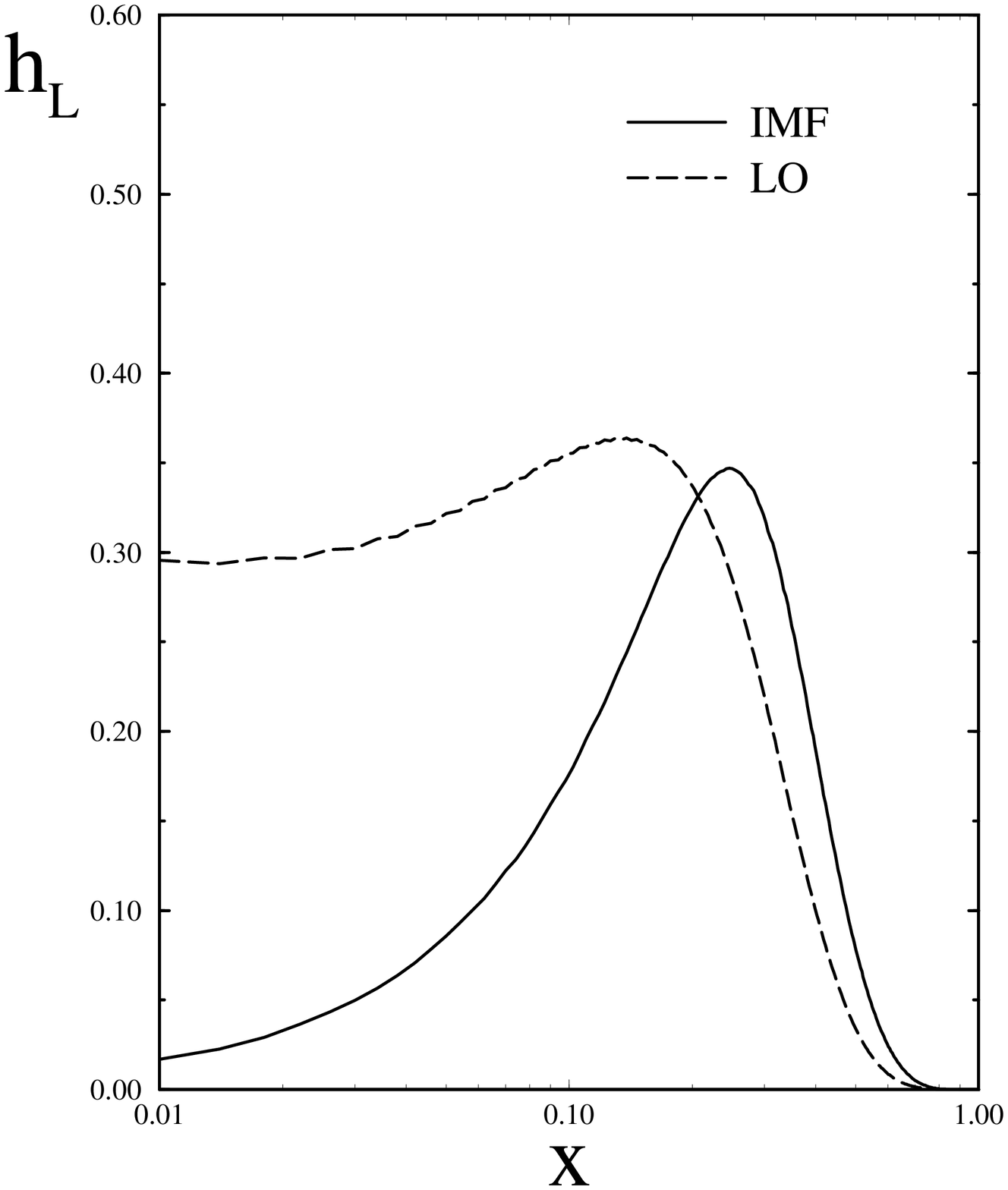,height=4.0cm,width=7.5cm}
\vspace{-0.4cm}
\caption{\label{fig_5}Same as fig \protect\ref{fig_4} 
for the chiral odd structure functions, evolved to 
$Q^2=4{\rm GeV}^2$.}
\vspace{-0.4cm}
\end{figure}

In figure \ref{fig_4} we display the analogous results for the 
polarized structure functions $g_1$ and $g_2$ which are obtained 
from the antisymmetric piece of the hadronic tensor. Details of the
calculations are presented in ref \cite{We97b}. Here we wish 
to mention that $g_2$ contains both twist--2 and twist--3 pieces
which are treated separately under the QCD evolution. We also stress
that the starting point $\mu^2=0.4{\rm GeV^2}$ of this evolution
is no longer a free parameter.
Apparently the model reproduces the empirical data quite well, 
although the associated error bars are sizable. Related quantities 
are the chiral odd structure functions $h_T(x)$ and $h_L(x)$. These 
may similarly be defined to the hadronic tensor (\ref{deften}), however,
as the correlation between a current and the scalar density 
${\bar \Psi}\Psi$. Eventually these structure functions will be 
obtained from the fragmentation region of DIS. The corresponding
NJL soliton calculation is reported in ref \cite{Ga98} and the 
results are shown in figure \ref{fig_5}.

\vspace{-0.2cm}
\section{EXTENSION TO FLAVOR SU(3)}
\vspace{-0.2cm}

\begin{figure}[ht]
\epsfig{figure=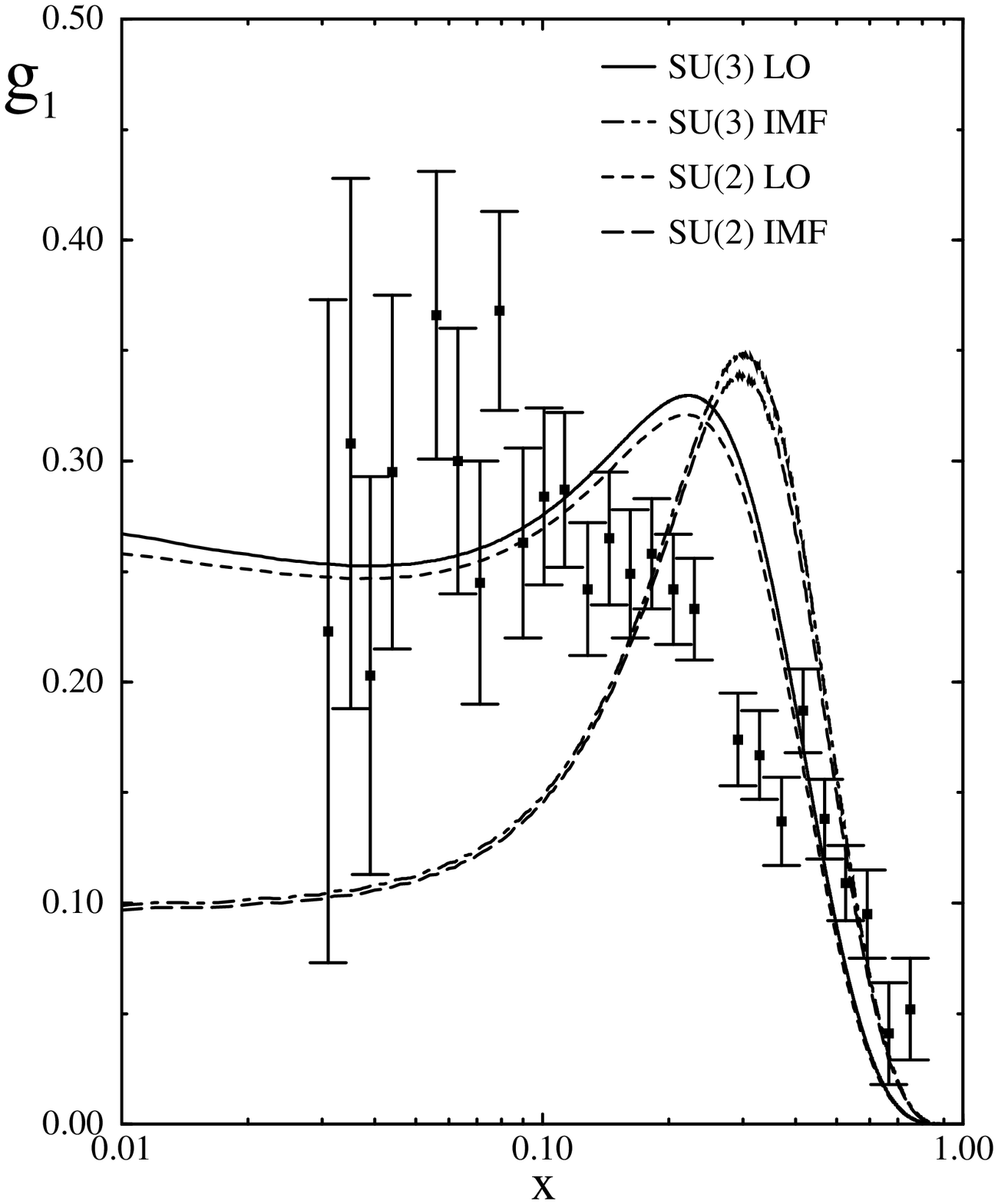,height=4.0cm,width=7.5cm}
~\vspace{0.3cm}\\
\epsfig{figure=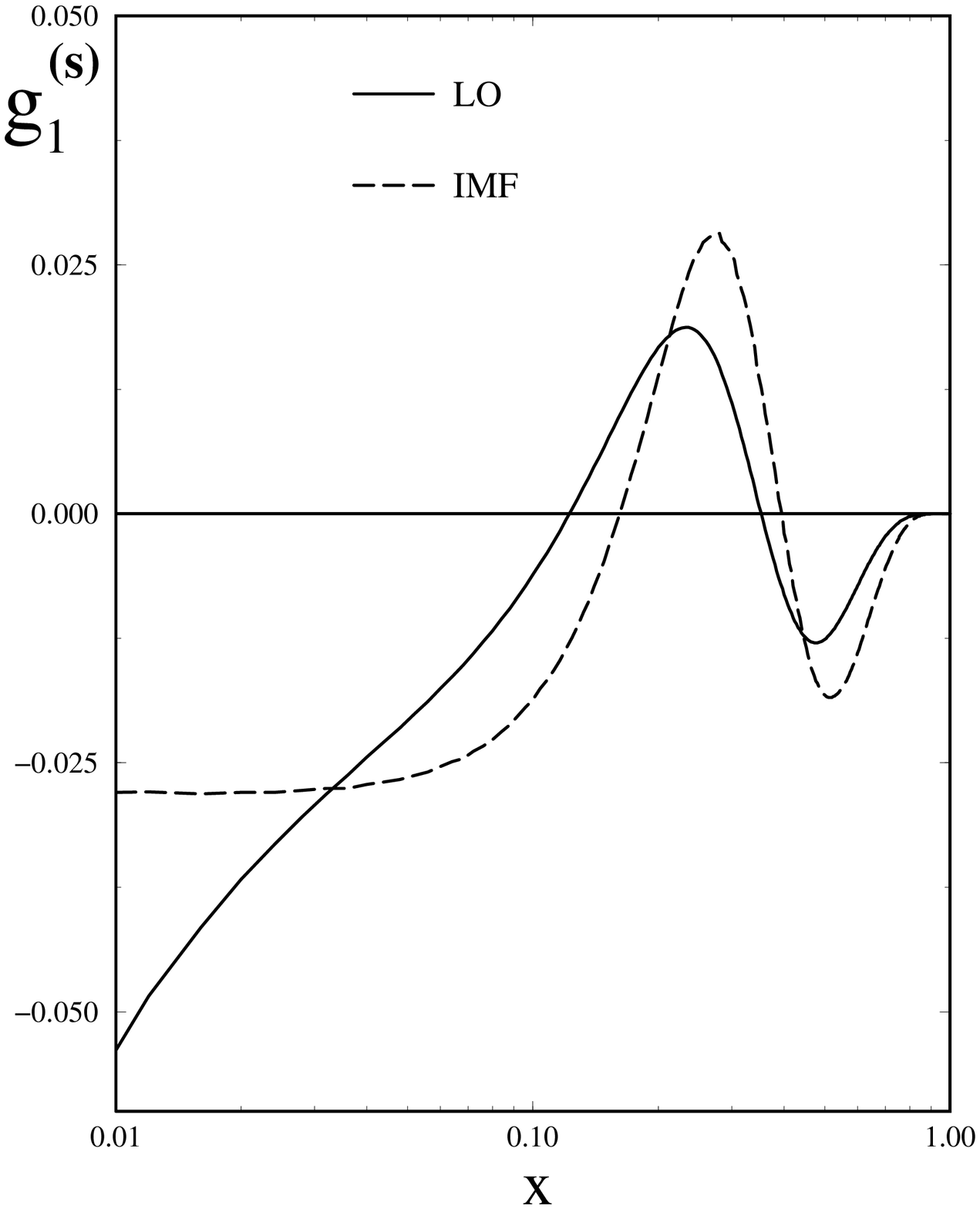,height=4.0cm,width=7.5cm}
\vspace{-0.4cm}
\caption{\label{fig_6}The polarized structure functions $g_1$
in the three flavor model. Top: Comparison between two 
and three flavor model. Bottom: Strangeness contribution.} 
\vspace{-0.4cm}
\end{figure}

The soliton picture has the celebrated feature that it 
can be extended to flavor SU(3) (for a review see \cite{We96a})
by generalizing to $A(t)\in{\rm SU(3)}$ in eq (\ref{collco}).
A detailed discussion of the three flavor NJL chiral soliton model is 
given in ref \cite{We92}. 
The resulting structure function $g_1(x)$ for the proton as 
calculated in \cite{Sch98} is shown in figure \ref{fig_6}.
Apparently the differences to flavor SU(2) are only minor 
and (after projection and evolution) the data are reasonably 
reproduced.
Moreover this model allows a projection to the strangeness 
contribution $g_1^{(s)}$. Also this is shown in figure \ref{fig_6}.
Apparently the smallness of the first moment of $g_1^{(s)}$ is 
due to cancellations between positive and negative pieces.

\vspace{-0.2cm}
\section{CONCLUSIONS}
\vspace{-0.2cm}

We have presented a calculation of nucleon structure functions 
within a chiral soliton model. We have argued that the soliton 
approach to the bosonized version of the NJL--model is most suitable 
since (formally) the required current operator is identical to the one 
in a free Dirac theory. Hence there is no need to approximate the current 
operator by {\it e.g.} performing a gradient expansion. Although the 
calculation contains a few (well--motivated) approximations it reproduces 
the gross features of the data after taking projection and evolution 
into account. This happens to be the case for both the polarized as well 
as the unpolarized structure functions.


\vspace{-0.2cm}
\begin{thebibliography}{9}
\vspace{-0.2cm}
\small
\bibitem{Na61}
Y. Nambu and G. Jona--Lasinio,
Phys. Rev. {\bf 122} (1961) 345; {\bf 124} (1961) 246.

\bibitem{Eb86}
D. Ebert and H. Reinhardt, 
Nucl. Phys. {\bf B271} (1986) 188.

\bibitem{Al96}
R. Alkofer, H. Reinhardt and H. Weigel,
Phys. Rep. {\bf 265} (1996) 139.

\bibitem{Ch96}
C. Christov, {\it et al.},
Prog. Part. Nucl. Phys. {\bf 37} (1996) 91.

\bibitem{We96}
H. Weigel, L. Gamberg and H. Reinhardt,
Mod. Phys. Lett. {\bf A11} (1996) 3021.

\bibitem{We97a}
H. Weigel, L. Gamberg and H. Reinhardt,
Phys. Lett. {\bf B399} (1997) 287.

\bibitem{We97b}
H. Weigel, L. Gamberg and H. Reinhardt,
Phys. Rev. {\bf D55} (1997) 6910.

\bibitem{Sch98}
O. Schr\"oder, H. Reinhardt and H. Weigel,\\
hep--ph/9805251.

\bibitem{Di96}
D. I. Diakonov {\it et al.},
Nucl. Phys. {\bf B480} (1996) 341;
M. Wakamatsu and T. Kubota,
Phys. Rev. {\bf D57} (1998) 5755.

\bibitem{Ja75}R. L. Jaffe,
Phys. Rev. {\bf D11} (1975) 1953;
R. L. Jaffe and A. Patrascioiu,
Phys. Rev. {\bf D12} (1975) 1314.

\bibitem{Ja80}
R. L. Jaffe,
\newblock Ann. Phys. (NY) {\bf 132} (1981) 32.

\bibitem{Ga97}
L. Gamberg, H. Reinhardt and H. Weigel,\\
hep--ph/9707352, Int. J. Mod. Phys. {\bf A}, in press.

\bibitem{Ar94}
M. Arneodo, {\it et al.}\ (NMC),
Phys. Rev. {\bf D50} (1994) R1.

\bibitem{Go67}
K. Gottfried,
Phys. Rev. Lett. {\bf 18} (1967) 1174.

\bibitem{slac96}
K.\ Abe {\em et al.},
Phys. Rev. Lett. {\bf 74} (1995) 346,
Phys. Rev. Lett. {\bf 75} (1995) 25,
Phys. Rev. Lett. {\bf 76} (1996) 587.

\bibitem{Ga98}
L. Gamberg, H. Reinhardt and H. Weigel,\\
hep--ph/9801379, Phys. Rev. {\bf D}, in press.

\bibitem{We96a}H. Weigel,
Int. J. Mod. Phys. {\bf A11} (1996) 2419.

\bibitem{We92}
H. Weigel, R. Alkofer and H. Reinhardt,
Nucl. Phys. {\bf B387} (1992) 638; A. Blotz {\it et al.}
Nucl. Phys. {\bf A555} (1993) 765.

\end{thebibliography}
\end{document}